\documentclass[conference]{IEEEtran}
\IEEEoverridecommandlockouts
\usepackage{cite}
\usepackage{amsmath,amssymb,amsfonts}
\usepackage{algorithmic}
\usepackage{graphicx}
\usepackage{longtable}
\usepackage{url}
\usepackage{textcomp}
\usepackage{xcolor}
\def\BibTeX{{\rm B\kern-.05em{\sc i\kern-.025em b}\kern-.08em
    T\kern-.1667em\lower.7ex\hbox{E}\kern-.125emX}}
\begin{document}

\title{Politics in Games - An Overview and Classification}

\author{
    \IEEEauthorblockN{Lisa Gutwenger\IEEEauthorrefmark{1},
                      Stephan Keller\IEEEauthorrefmark{2},
                      Martin Dolezal\IEEEauthorrefmark{1},
                      Bernhard Schnögl\IEEEauthorrefmark{2},
                      Sebastian Rous\IEEEauthorrefmark{1},
                      Klaus Poier\IEEEauthorrefmark{1},
                      Johanna Pirker\IEEEauthorrefmark{2}\IEEEauthorrefmark{3}}
    \IEEEauthorblockA{\IEEEauthorrefmark{1}University of Graz, Austria}
    \IEEEauthorblockA{\IEEEauthorrefmark{2}Graz University of Technology, Austria}
    \IEEEauthorblockA{\IEEEauthorrefmark{3}Ludwig Maximilian University of Munich, Germany}
}

\maketitle

\begin{abstract}
The representation of politics in media influences societal perceptions and attitudes. Video games, as a pervasive form of media, contribute significantly to this phenomenon. In this work, we explore political themes within video games by analyzing politically-themed games on game distribution platforms including Steam. We conducted a statistical examination of games with political context to identify patterns and use this as a basis to introduce a first taxonomy to categorize and better understand the interplay between politics and video games. This taxonomy offers a first framework for analyzing political content in games and also sets a foundation for future research in this field.
\end{abstract}


\section{Introduction}
Media, especially social media in recent years, has significantly impacted political opinions. Many previous publications have examined the influence of platforms like Twitter and Facebook on opinion formation \cite{kruse2018social}. One medium sometimes overlooked in these analyses is video games. However, an increasing number of studies demonstrate their impact on opinion formation and address issues like radicalization through games \cite{lopez2024video}. In this work, we want to understand the representation of politics in video games on the Steam platform and build an overview of future research directions. 

This research intersects various interdisciplinary domains, aiming to unravel how politics is portrayed in video games. Our study is threefold: examining player demographics to identify the audience, exploring game characteristics to understand political representations, and investigating the potential consequences of such representations in video games. Following, we discuss characterization categories relevant to understanding the context of politics in video games.

\section{Discussing Games}

To discuss games and understand the games we will use different categories: player demographics, categorization, age recommendation, and possible consequences.  

\textbf{Player Demographics}. A comprehensive understanding of the audience is fundamental. According to the Entertainment Software Association\cite{esa2022} (2022), 65\% of adults and 71\% of minors in the US are gamers and this number is growing. Also, over the past two decades, female participation has risen, reaching near parity. However, the age distribution has remained stable, with the average gamer being 33 years old. These demographic shifts are pivotal in contextualizing the reception and impact of political content in games.

\textbf{Categorization of Games} 
There are different ways to categorize games. We can categorize games based on their primary purpose (entertainment vs. serious gaming) and their persuasive intent \cite{bopp2009,raessend2019}. Another categorization is game genres \cite{breiner2019} based on the in-game activities. The popularity of genres varies regionally, with strategy, adventure, and simulation leading in the DACH region, contrasting with the US preferences\cite{statista2022-usa, statista2022-de}. We can also categorize by political dimension. This can be described by analyzing the representation of governance systems in games such as autocracy, democracy, or a selectable system. Examples include Sid Meier's Civilization (selectable system), or Papers, Please! and Beholder (autocratic themes).

\textbf{Age Recommendations}
Age ratings (ESRB, PEGI, USK) are vital in understanding the intended audience. However, these systems are not without criticism\cite{duffy2022}, raising questions about their efficacy and relevance in the context of political content and potential educational experiences.

\textbf{Possible Consequences}
The impact of gaming on younger generations has been a contentious topic, often skewed towards negative consequences such as isolation and aggression, culminating in the WHO's classification of "gaming disorder" in ICD-11. Nevertheless, this perspective is increasingly challenged\cite{darvesh2020}. Research is now focusing on specific effects, including the influence of games on political attitudes and behaviors. Fewer studies address this compared to traditional effects like aggression. Bacovsky\cite{bacovsky2021a} investigates the political engagement of gamers, while Dill et al.\cite{dill2008} and Saleem \& Anderson\cite{saleem2013} explore the reinforcement of gender stereotypes and racial prejudices in games. These studies highlight the multifaceted impact of video games, extending beyond mere entertainment to shaping societal views and behaviors.

To understand better games with a focus on political context, we can describe different categories: 
\begin{itemize}
    \item Based on activity (e.g. simulation, strategy, action)
    \item Based on purpose/intend (e.g. education, understanding, assessment, persuasion, entertainment) 
    \item Based on the player perspective (e.g. leader of a country, civilian, neutral) 
    \item Based on the historical context, setting 
    \item The representation form of the government (e.g. democracy, autocracy,..) 
\end{itemize}

Based on these categories, we want to understand games with political context published on game distribution platforms. 

\section{Analysis}

This study focuses on the representation of politics in video games. We initially identified a sample of politically-themed games from the Steam and Microsoft Store platforms using the keywords "politic", "democrats", "republicans", "democracy", and "dictatorship". After a manual refinement process by the authors to exclude irrelevant entries and the inclusion of games from relevant literature, our analysis encompassed 28 games. This collection of political-themed video games offers a diverse exploration of global and historical political scenarios through various gameplay styles. In table \ref{table:overview} a list of the analyzed games is given and following we discuss a few examples in more detail. 

\textit{"Realpolitiks New Power"} 
 immerses players in modern geopolitical challenges, encouraging them to navigate complex international affairs. \textit{"Dictators: No Peace Countryballs"} 
and \textit{"Tropico 6"} 
put players in the shoes of a dictator, blending strategy and simulation to shape national and global politics. \textit{"Supreme Ruler: Ultimate"} 
spans historical and futuristic contexts, allowing players to lead any nation through tumultuous times, while \textit{"Suzerain"} 
offers a narrative-driven experience as a president amidst political drama.
\textit{"Democracy 4"} 
and \textit{"Government Simulator"} 
 offer simulations of governing, with the former focusing on policy-making and re-election, and the latter on managing a country's various aspects based on real data. \textit{"World Peace General 2017"} 
provides a strategic challenge of world domination or destruction.
\textit{"Power \& Revolution"} 
simulates being a state head with up-to-date global issues, including the COVID-19 pandemic, and \textit{"Rebel Inc: Escalation"} 
tasks players with rebuilding a region post-conflict. \textit{"China: Mao's Legacy"} 
delves into historical political drama in post-Cultural Revolution China.
Other games like \textit{"Smart City Plan"} 
and \textit{"Workers \& Resources: Soviet Republic"} 
focus on city-building and management, integrating political decision-making. \textit{"The Political Process"} 
offers an in-depth simulation of the American political system, where players can rise through the ranks of political office.

In our analysis of the representation of politics in video games, we observed a diverse array of genres with a predominant focus on simulation \textit{and strategy}. This categorization aligns with the intricate nature of political scenarios that these games aim to simulate. Specifically, genres such as simulation games and strategy were most prevalent. These genres are well-suited to convey the complexities and nuances of political decision-making and governance. Also Text-Adventure/RPG was a relevant genre with "Suzerain" being an example.


The data also indicated a range of player perspectives, primarily centering around leadership roles. Titles like "Realpolitiks New Power"
, "Tropico 6" 
place players in the position of a country's leader, often as a dictator or president. This choice of perspective is critical in immersing players in the responsibilities and challenges inherent in political leadership.

Regarding the price range, the games varied significantly, with some being as economical as 0.79€, while others like "Tropico 6" were priced up to 39.99€. This variation in pricing points to the accessibility of these games for a wide spectrum of players. Additionally, the presence of free titles in our dataset further underscores the inclusivity of this genre.

The historical context of these analyzed games ranged from modern world issues, as seen in "Realpolitiks New Power," to historical and futuristic scenarios in "Supreme Ruler: Ultimate."
This diversity in historical settings suggests that political video games are not only a medium for entertainment but also for education and reflection on both historical and contemporary political issues.
Moreover, the availability of these games on popular platforms such as Steam and Microsoft Store indicates their widespread accessibility, catering to a broad audience with varying interests in political themes.

In summary, our analysis underscores the significant role of video games as a medium for exploring and understanding political themes. By offering a range of genres, player perspectives, and historical contexts, these games provide valuable insights into the dynamics of political systems and decision-making processes, thereby serving as a unique blend of entertainment and educational tool.

\subsection{Classification Criteria}
The games were systematically categorized based on multiple criteria:
\begin{itemize}
\item Developer studio and publisher details, including the instructing party in case of contract development.
\item Release year to trace historical trends.
\item Game genre and its classification as either an entertainment or serious game.
\item Age recommendation, based on ESRB, PEGI, and USK ratings.
\item Nature of the game (commercial or free).
\item Content and platform of the game.
\item Representation of the form of government.
\end{itemize}

\subsubsection*{Prevalent Regime Classification}
Initial classification results reveal a balanced distribution of political themes. Approximately 39\% of the games focus on democratic systems, another 39\% on autocratic systems, and the remaining 22\% allow players to choosye the government form.

\subsubsection*{Age Recommendations Analysis}
When aggregating the data from all three rating systems, the average recommended age is approximately 11.3 years (ESRB = 12.2 years, PEGI = 9.8 years, USK = 12.2 years).

\subsection{Steam Dataset Analysis}
\begin{figure}
\centering
\includegraphics[width=\linewidth]{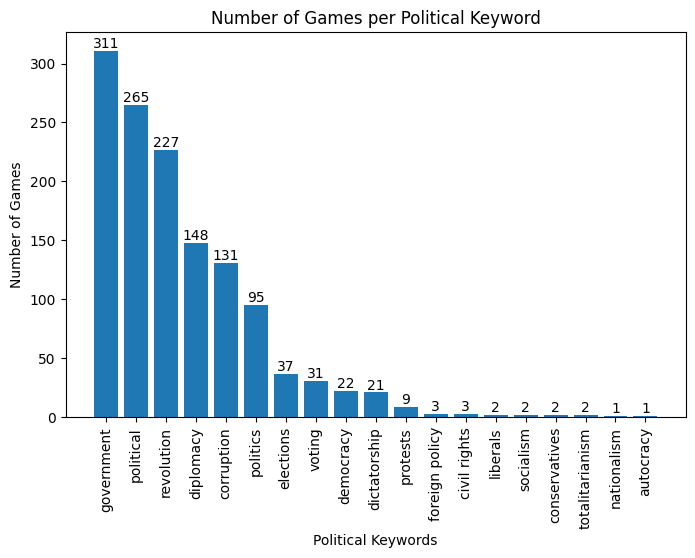}
\caption{Number of games containing each keyword (out of n=27,000).}
\label{fig:number_total}
\end{figure}

To broaden our research scope, we utilized a dataset from Kaggle containing information on 27,000 Steam games\cite{davis2019}. Due to the dataset's size, we relied on automated keyword searches rather than manual evaluation. We found that only 0.16\% (44 out of 27,000) of the games included relevant keywords such as "democracy", "dictatorship", or "autocracy", with "autocracy" appearing only once. Expanding our keyword list provided further insights into the prevalence of political themes in games (Figure \ref{fig:number_total}).

\begin{figure}
\centering
\includegraphics[width=\linewidth]{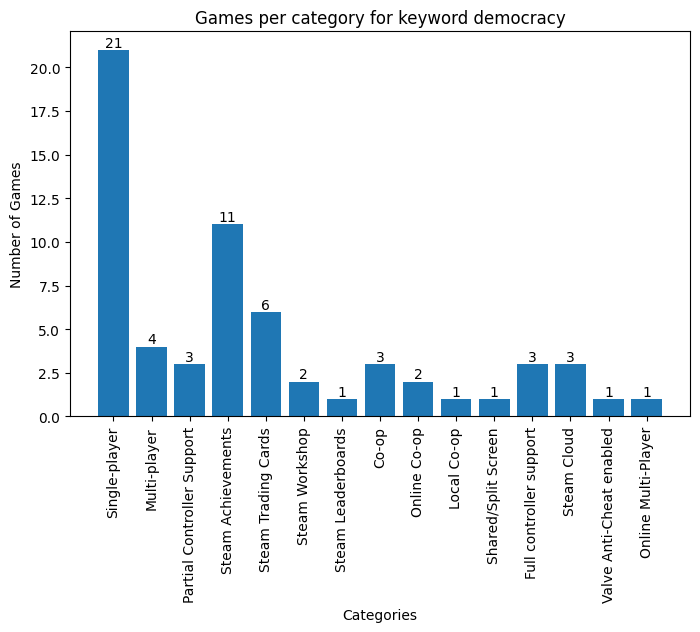}
\caption{Distribution of the keyword "democracy" across game categories.}
\label{fig:steam_categories}
\end{figure}


\section{Outlook}

This research has laid the groundwork for a nuanced understanding of political themes in video games and to understand better classification options. We observed a prevalence of political themes in video games, which are represented in a variety of forms. This diversity in representation opens questions about the influence of these games on societal and political perceptions, particularly among younger demographics. We can see that these games have the potential to shape the political understanding and beliefs of their players, making this an area of interest for further exploration. Furthermore, our analysis revealed a potential correlation between the genres of these games and their political content. This suggests that different types of games may engage with political themes in diverse ways, which could have profound implications for game design and the study of political socialization. The way these games are designed and the political narratives they choose to convey can potentially influence players' understanding and engagement with political concepts.

Overall, this work presents a first overview of games and politics context to support our understanding of the intersection between politics and video games but also opens up numerous possibilities for future research that can bridge the gap between game studies and political science. The initial categorization of games based on political themes shows promise. However, there's a need for finer granularity in categorizing political aspects to capture the subtle nuances in different games. This refinement will enable more precise and meaningful analysis.

\vspace{12pt}

\onecolumn

\begin{longtable}[c]{|p{4cm}|p{1.5cm}|p{0.6cm}|p{3cm}|p{2.0cm}|p{3.5cm}|}
\hline
\textbf{Game Name} & \textbf{Gov} & \textbf{Year} & \textbf{Genre} & \textbf{Purpose} & \textbf{Topic} \\
\hline
\endfirsthead
\multicolumn{6}{c}%
{{\bfseries Table \thetable\ continued from previous page}} \\
\hline
\textbf{Game Name} & \textbf{Government/Leadership Form} & \textbf{Year} & \textbf{Genre} & \textbf{Purpose} & \textbf{Topic} \\
\hline
\endhead
Bad News & Democracy & 2018 & Adventure & Education & Disinformation \\
Beholder & Autocracy & 2016 & Adventure, Strategy & Empathy, Understanding & Regime, Privacy \\
Command \& Conquer (RC) & Autocracy & 2020 & Strategy & Entertainment & Military RTS \\
Crusader Kings III & Autocracy & 2020 & Role-Playing, Simulation, Strategy & Education, Entertainment & Historical Accuracy \\
Democracy: The Isle of Five & Democracy & 2020 & Simulation, Strategy & Discussion, Education & Cooperation vs. Competition, Empathy \\
Democracy 4 & Democracy & 2022 & Simulation, Strategy & Entertainment & Political Action, Education Potential \\
Dictators: No Peace Countryballs & Autocracy & 2020 & Simulation, Strategy & Entertainment & Dictator Simulation \\
Dope Elections & Democracy & 2020 & Platform & Political Education & Support for Elections in CH \\
Europa Universalis IV & Autocracy & 2013 & Simulation, Strategy & Entertainment & Historical Accuracy \\
Frostpunk & Autocracy & 2018 & Simulation, Strategy & Entertainment & Decisions about Systems \\
Government Simulator & Democracy & 2017 & Simulation, Strategy & Serious Game & Real World Data, Simulation \\
Leon's Identity & Democracy & 2020 & Adventure & Extremism Prevention & Education about Radicalization \\
Power and Revolution & Selectable & 2016 & Simulation, Strategy & Serious Game & Geopolitical Simulator \\
Realpolitiks - New Power & Selectable & 2018 & Strategy & Entertainment & RTS for Macromanagement \\
Rebel Inc: Escalation & Democracy & 2021 & Casual Simulation, Strategy & Education, Awareness & Counterinsurgency \\
Rogue State & Autocracy & 2015 & Simulation, Strategy & Entertainment & Infrastructure and Building Simulation \\
Sid Meier's Civilization VI & Selectable & 2016 & Strategy & Entertainment, Education & Historical Accuracy \\
Smart City Plan & Democracy & 2020 & Simulation, Strategy & Entertainment & City Planning \\
Superpower 2 & Selectable & 2004 & Simulation, Strategy & Entertainment & Strategy Simulation \\
Supreme Ruler Ultimate & Selectable & 2014 & Simulation, Strategy & Entertainment & Strategy Simulation \\
Suzerain & Democracy & 2020 & Adventure, Role-Playing & Serious Games & Political Decisions as a President \\
The Political Process & Democracy & 2019 & Simulation, Strategy & Education & Politics Simulation \\
Total War: Three Kingdoms & Autocracy & 2019 & Action Strategy & Entertainment & World Simulation \\
Tropico 6 & Autocracy & 2019 & Simulation, Strategy & Education & Democracy Simulator \\
Urban Empire & Selectable & 2017 & Simulation, Strategy & Entertainment & Strategy Decisions as a Ruler \\
Voting Rush Election Game & Democracy & 2016 & Adventure & Entertainment & Sorting Game \\
World Peace General 2017 & Autocracy & 2017 & Strategy & Entertainment & Resource Management \\
Papers, Please & Autocracy & 2013 & Adventure, Puzzle & Education, Awareness & Moral Dilemma, Decisions, Immigration \\
\hline
\caption{Summary of Games with their Government Forms, Genres, and Purposes}
\label{table:overview}
\end{longtable}
\twocolumn

\end{document}